\documentclass{emulateapj}
\usepackage{epsfig}
\usepackage{amsmath}
\usepackage[dvips]{color}

\begin{document}

\def\Ecut{E_{\rm cut}}
\def\C{{\mathcal C}}

\sloppy

\def\be{\begin{equation}}
\def\ee{\end{equation}}
\def\ba{\begin{eqnarray}}
\def\ea{\end{eqnarray}}
\def\lsim{\raise0.3ex\hbox{$\;<$\kern-0.75em\raise-1.1ex\hbox{$\sim\;$}}}
\def\gsim{\raise0.3ex\hbox{$\;>$\kern-0.75em\raise-1.1ex\hbox{$\sim\;$}}}
\def\ap{\approx}
\def\eps{\varepsilon}
\def\theta{\vartheta}

\def\alt{\raise0.3ex\hbox{$\;<$\kern-0.75em\raise-1.1ex\hbox{$\sim\;$}}}
\def\agt{\raise0.3ex\hbox{$\;>$\kern-0.75em\raise-1.1ex\hbox{$\sim\;$}}}

{\hfill\\{}\\ FERMILAB-PUB-08-315-A, IFT-UAM/CSIC-08-55}

\title{A global autocorrelation study after the first Auger data: impact on the number density of UHECR sources}

\author{A.~Cuoco$^{1}$, S.~Hannestad$^{1}$, T.~Haugb{\o}lle$^{1}$, M.~Kachelrie\ss$^{2}$, P.~D.~Serpico$^{3,4}$}

\affil{$^1$ Department of Physics and Astronomy, University of
Aarhus, Ny Munkegade, Bygn. 1520, DK--8000 Aarhus, Denmark}

\affil{$^2$ Institutt for fysikk, NTNU, N--7491 Trondheim, Norway}

\affil{$^3$ Center for Particle Astrophysics, Fermi National
Accelerator Laboratory, Batavia, IL 60510-0500, USA}

\affil{$^4$ Physics Division, Theory Group, CERN, CH-1211 Geneva 23,
Switzerland}
%\date{\today}

\begin{abstract}
We perform an autocorrelation study of the Auger data with
the aim to constrain the number density $n_s$  of ultrahigh energy
cosmic ray (UHECR) sources, estimating at the same time the effect
on $n_s$ of the systematic energy scale uncertainty and of the distribution
of UHECR. The use of
global analysis has the advantage that no biases are introduced,
either in $n_s$ or in the related error bar, by the \emph{a priori}
choice of a single angular scale. The case of continuous, uniformly
distributed sources is nominally disfavored at 99\% C.L. and the fit improves
if the sources follow the large-scale structure of matter in the
universe. The best fit values obtained for the number density of
proton sources are within a factor $\sim$2 around $n_s\simeq 1\times 10^{-4}/$Mpc$^3$
and depend mainly on the Auger energy calibration scale, with lower densities being
preferred if the current scale is correct. The data show no
significant small-scale clustering on scales smaller than a few degrees. This might be interpreted as a signature of
magnetic smearing of comparable size, comparable with the indication
of a $\ap 3^\circ$ magnetic deflection coming from cross-correlation
results. The effects on the above results of some approximations done is also discussed.
\end{abstract}


%\pacs{98.70.Sa  %Cosmic rays (including sources, origin, acceleration, and
                %interactions)}
                \maketitle

%%%%%%%%%%%%%%%%%%%%%%%%%%%%%%%%%%%%%%%%%%%%%%%%%%%%%%%%%%%%%%%%%%%%%%
\section{Introduction}\label{intro}
%%%%%%%%%%%%%%%%%%%%%%%%%%%%%%%%%%%%%%%%%%%%%%%%%%%%%%%%%%%%%%%%%%%%%%
Evidence is now emerging that ultrahigh energy cosmic rays (UHECRs)
have an astrophysical origin, as opposed to being generated in
exotic top-down models: The detection of a spectral suppression
consistent with the Greisen-Zatsepin-Kuzmin (GZK) effect
\cite{Greisen:1966jv,Zatsepin:1966jv} by both
HiRes~\cite{Abbasi:2007sv} and Auger~\cite{Abraham:2008ru}
collaborations, together with the Auger bounds on the UHE neutrino
flux~\cite{Abraham:2007rj}, on the photon
fraction~\cite{Aglietta:2007yx} and on the anisotropy towards the
Galactic Center~\cite{Aglietta:2006ur,Aloisio:2007bh} are all
consistent with this scenario. The next step is clearly the
identification of the sources of UHECRs, an arena where anisotropy
studies play a crucial role. Yet, the limited angular resolution of
extensive air shower detectors and especially the deflections that
charged particles suffer in astrophysical magnetic fields make the
task highly non trivial. This is especially troublesome given that
the UHECRs chemical composition is unknown, that we lack a detailed
knowledge of the Galactic magnetic field structure and, above all,
of the very magnitude and structure of extragalactic magnetic fields
(EGMF) outside of cluster cores. These limitations---together with
the small statistics available at present---suggest that, at least
in an initial phase, charged particle astronomy may be limited to
the inference on the statistical properties of UHECR sources, rather
than a detailed study of single accelerators.

In Ref.~\cite{I}, we found that a global comparison of the
two-point auto-correlation function of the data with the one of
catalogues of potential sources is a powerful diagnostic tool: This
observable is less sensitive to unknown deflections in magnetic fields
than the cross-correlation function, while keeping a strong discriminating
power among source candidates. In particular,  the autocorrelation
function of (sub-) classes of galaxies have different biases with
respect to the large-scale structure (LSS) of matter. As a result,
the best fit value for the density $n_s$  of different source
classes may differ, especially if only one or a small range of
angular scales is considered. Although the bias of different source
classes differs maximally at small angular scales, we showed that
the statistically most significant differences are at intermediate angular
scales, where both the larger number of cosmic ray pairs (CR) and of
galaxy pairs leads to relatively smaller error bars. Moreover, the
autocorrelation function on larger angular scales becomes less
dependent on possible deflections in the Galactic and extragalactic
fields.\\
\indent
In this article we derive the number density of UHECR sources using
the recently published arrival directions and energies of the 27
Auger events~\cite{PAO} with estimated energy $E\geq 57$\,EeV,
thereby complementing the study~\cite{I} with a concrete example for
a comparison of the {\em global\/} cumulative autocorrelation
function of sources and UHECRs. Note that we showed in Ref.~\cite{I} that,
even in an idealized case where systematics play no major role,
roughly three times the number of ``useful''
events that can be extracted from Ref.~\cite{PAO} are required to start
distinguishing between different sub-classes of sources. Thus a
study of the kind envisaged in Ref.~\cite{I} is unrealistic at present.
Here, we restrict ourselves to more modest goals: i) To compare
the data against predictions of two toy model cases of uniformly
distributed sources and of ``normal galaxies''  (i.e.\ sources that
have the same distribution as the PSCz
catalogue~\cite{Saunders:2000af}) which we shall refer to with the
two values for the label $\kappa=\{{\rm uni},{\rm LSS}\}$,
respectively. ii) To study the effect on the allowed range of $n_s$
of a systematic error on the energy scale of the UHECR experiment.
Note that preliminary results of the clustering of the Auger events has been
presented in \cite{Mollerach:2007vb}, but astrophysical
implications have not been discussed there.

We review the statistical method we use in
Sec.~\ref{analysis}, and apply it in  Sec.~\ref{int} to the
Auger data, providing some interpretation of the results. In Sec.~\ref{disc} we discuss
some limitations and caveats of the analysis. Finally, we summarize in Sec.~\ref{sum}.
%%%%%%%%%%%%%%%%%%%%%%%%%%%%%%%%%%%%%%%%%%%%%%%%%%%%%%%%%%%%%%%%%%%%%%%%%%%%
\section{Statistical analysis of the data}\label{analysis}
%%%%%%%%%%%%%%%%%%%%%%%%%%%%%%%%%%%%%%%%%%%%%%%%%%%%%%%%%%%%%%%%%%%%%%%%%%%%
The use of correlation functions is well suited to the study of
over- and underdensities of non-uniformly distributed sources
and of the resulting anisotropies in the radiation received from them.
Since the number of CR events published by Auger is still small, we
use in our analysis following Ref.~\cite{point} the
{\em cumulative\/} two-point autocorrelation function $\C(\theta)$
defined as
\be
 \C(\theta) = \sum_{i=2}^N\sum_{j=1}^{i-1}\Theta(\theta-\theta_{ij}) \,,
\ee
i.e.\ the  number of pairs within the angular distance $\theta$.
Here, $N$ is the number of CRs considered, $\theta_{ij}$ is the angular
distance between events $i$ and $j$ and $\Theta$ is the step function
(with $\Theta(0)=1$).

This function is straightforward to compute for the data, and
denoted then by $\C_\ast(\theta)$. For a specific model hypothesis
$X$, a set of  functions $\C_i(\theta|X)$ is obtained  in the
following way: Sources with equal luminosities are  distributed
within a sphere of $180h^{-1}$\,Mpc either uniformly or following
the three-dimensional LSS as given by a smoothed
version~\cite{Cuoco:2005yd,Cuoco:2007um} of the PSCz
catalogue~\cite{Saunders:2000af}. For the LSS case
(but not for the uniform case) sources and CR events within the
PSCz mask are excluded, leaving 22 CR events. Note that the mask
mostly overlaps with the Galactic plane and bulge region, where
larger deflections due to the Galactic magnetic field are expected:
The mask is thus not only a catalogue limitation, but also
implements to some extent a physically motivated angular cut.
Finally, each source $k$, at redshift $z_k$, is weighted by the factor
\be \label{weight}
 \frac{1}{z_k^2}  \int_{E_i(E_{\rm cut},z_k)}^{\infty}
  \!\!\!\!\!\!\!\! E^{-s} dE
\ee
accounting for its redshift dependent flux suppression and the CR
energy losses. These are parametrized as a continuous process in
term of the function $E_i(E_{f},z_k)$ which gives the initial
injection energy $E_i$ for a particle leaving the source at
$z_k$ and arriving at the Earth with energy $E_f<E_i$. Further
details are given in~\cite{Cuoco:2005yd}. The injection spectral
index $s$ is assumed to be the same for all the sources and equal to
2.0. The dependence on $s$ is however weak as shown in more detail
in~\cite{Cuoco:2005yd}. This procedure defines the model, while a
single random realization is obtained by choosing the observed
number of events from the
sources according to the source weights and the
declination-dependent Auger experimental exposure.

The model thus depends directly only on $n_s$ and the choice between
sources distributed uniformly or according to the PSCz catalogue (of course,
implicitly it also depends on the assumptions of sufficiently small magnetic field deflections).
Additionally, the model depends via the weights of Eq.~(\ref{weight}) on the
type of primary particle, the energy spectrum of the sources, and
the energy scale and cut $E_{\rm cut}$. The latter dependence
arises, because the energy scale of CR experiments has a relatively
large systematic error that is difficult to determine. In
particular, it has been argued~\cite{MT} that  the energy scale of
Auger should be shifted up by 30--40\% to obtain agreement with the
spectral shape predicted by $e^+e^-$ pair production~\cite{dip} and
the CR flux measured by other experiments. Therefore we use two
different values for the energy cut, $E_{\rm cut}=60$\,EeV assuming
that the energy calibration of Auger is correct and $E_{\rm
cut}=80$\,EeV inspired by the dip interpretation. We do not include
in this work the finite energy resolution of the Auger experiment
that is of order 20\% in $\Delta E/E$. A finite energy resolution
would result in an effective decrease of the nominal energy cut due
to the steeply falling CR spectrum and to a larger GZK
horizon~\cite{res}. Similarly, we do not account for the
stochasticity of the energy loss in the photo-pion production
process. Both of the effects are subdominant at the moment with the
present low statistics while a more careful analysis will be needed
in the future when more data is available. For a rough estimate of the influence of
both effects on $n_s$ one may compare how the 30\% up-ward shift of
$E_{\rm cut}$ from 60 to 80\,EeV changes $n_s$. Finally, throughout
this work we consider proton primaries, but note that the
combination of nuclei with large deflections and few sources has
been advocated too~\cite{nuclei,fargion,gorbunov08}. A further brief
discussion of this point is postponed to Sec.~\ref{chemadd}.

The cumulative autocorrelation of the data $\C_\ast(\theta)$ , which
is a single, one-dimensional function, has now to be compared with
the hypothesis $X$, for which we have  various Monte Carlo
realizations $\C_k(\theta|X)$, $k=1,\ldots,M,$ (we use typically
$M\sim10^5$). A standard method to compare data and model is to use
angular bins $\theta_i$ so that to substitute the continuous
function $\C(\theta)$ with the discrete set of values
$\C(\theta_i)$. The Monte Carlo realizations can then be used to
calculate the marginalized probability distribution of each single
$\C(\theta_i|X)$ or, if required, the joint probability distribution
of two $\C$ variables $\C(\theta_i|X), \C(\theta_j|X)$, three $\C$
variables or more. In practice, to derive the best fit value $n_s$
as well as the goodness-of-fit for the chosen hypothesis $X$ a
possible way is to calculate the mean $\langle\C(\theta_i|X)\rangle$
and the variance $\sigma_i$ per bin as well as the correlation
matrix $\sigma_{ij}$ and then to perform a $\chi^2$ test. However,
the difficulty to deal with such an high dimensional probability
space and the generally strong non-gaussianity of the probability
distribution make the $\chi^2$ method clearly not optimal for the
problem at hand.

\begin{figure*}[!tbp]
\begin{center}
\begin{tabular}{cc}
\epsfig{file=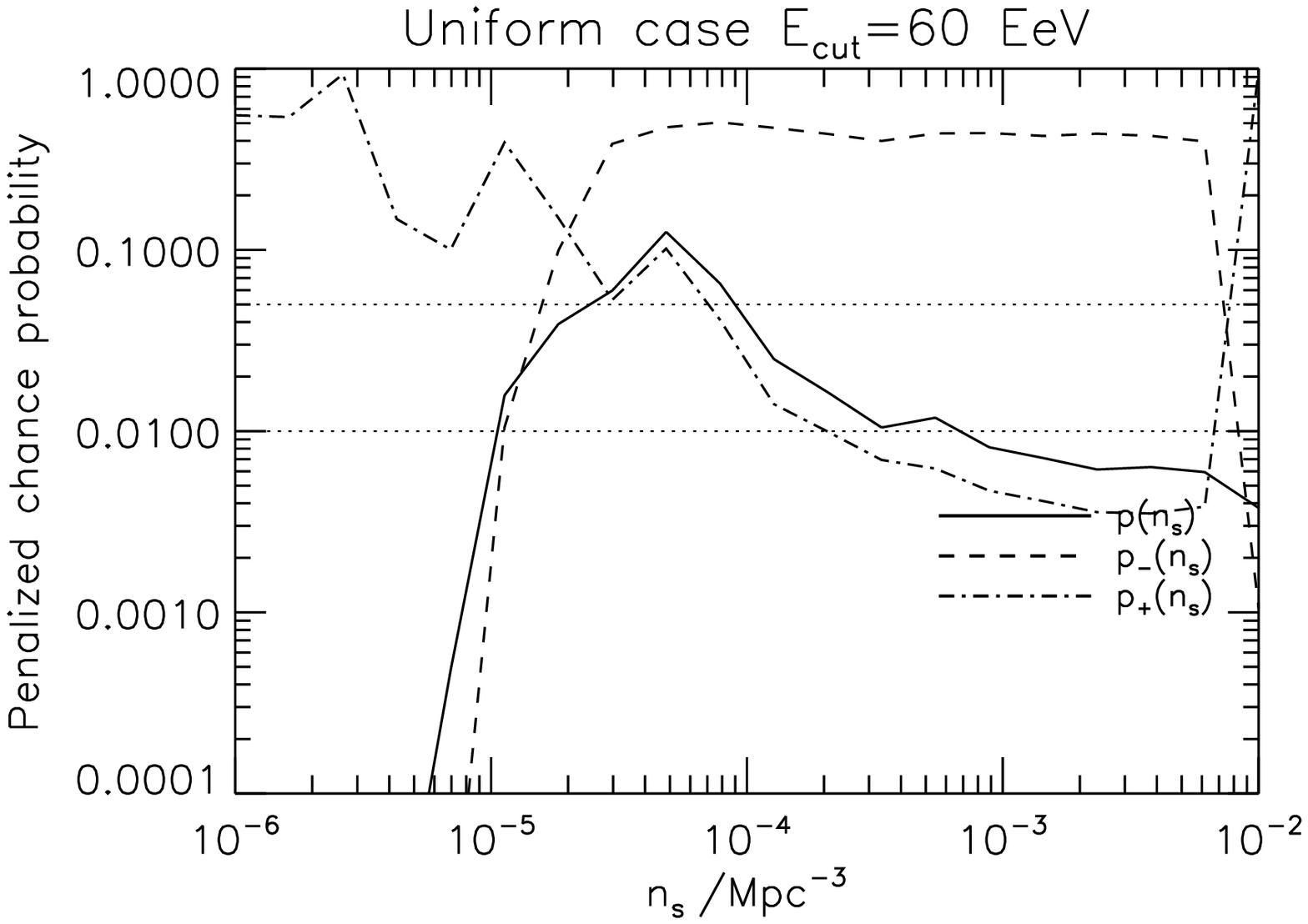,width=0.47\textwidth} &
\epsfig{file=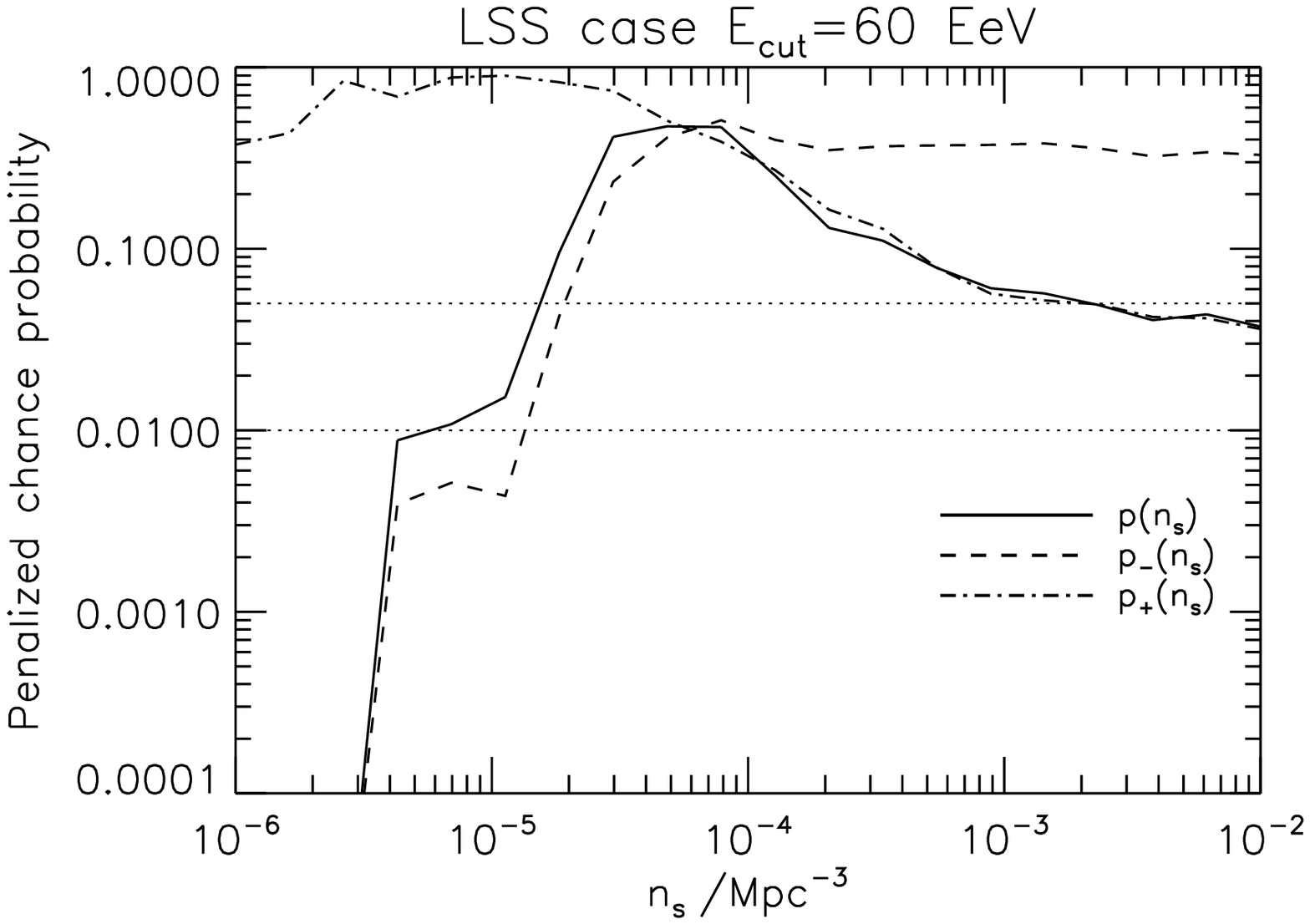,width=0.47\textwidth}\\
\end{tabular}
\begin{tabular}{cc}
\epsfig{file=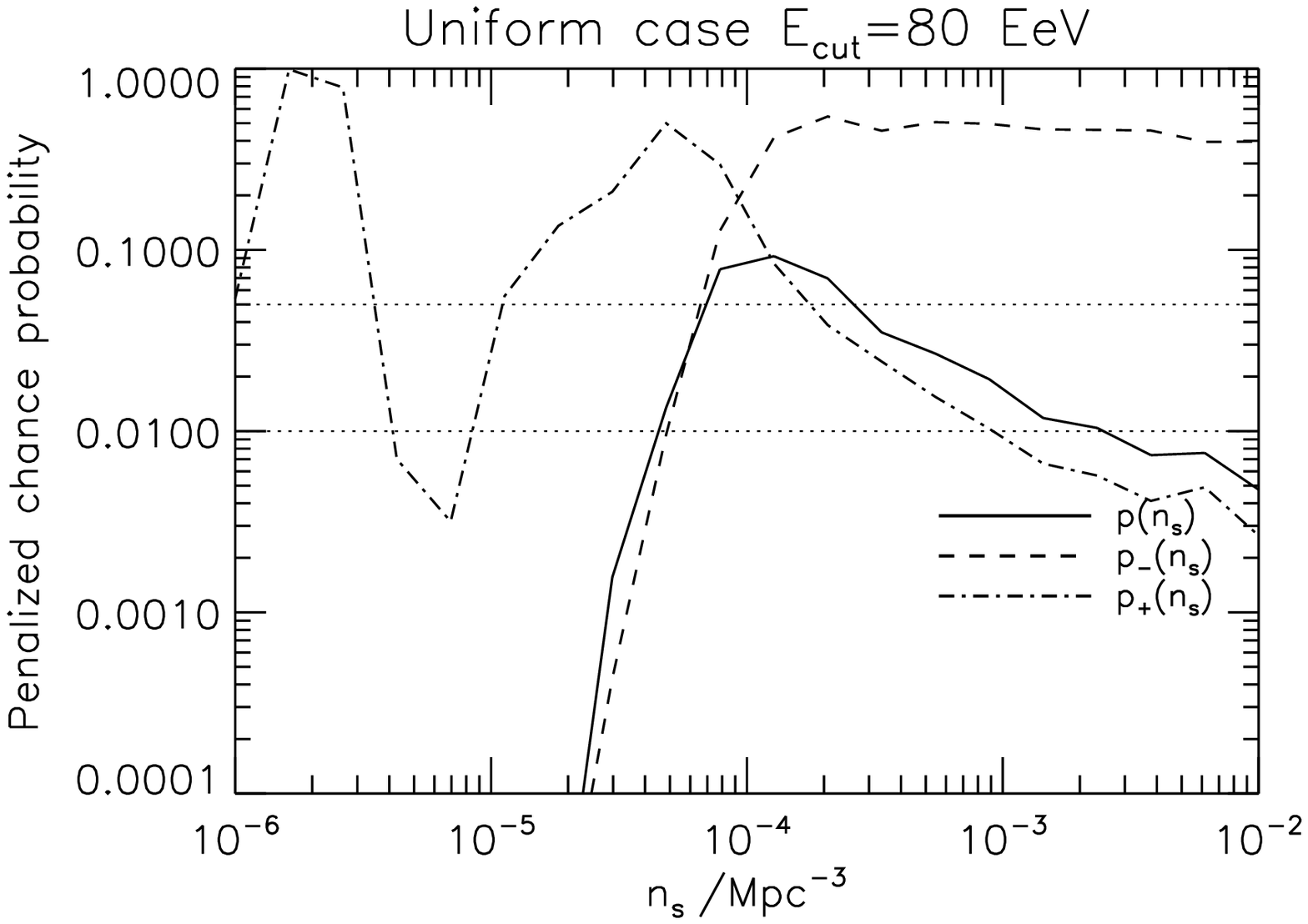,width=0.47\textwidth} &
\epsfig{file=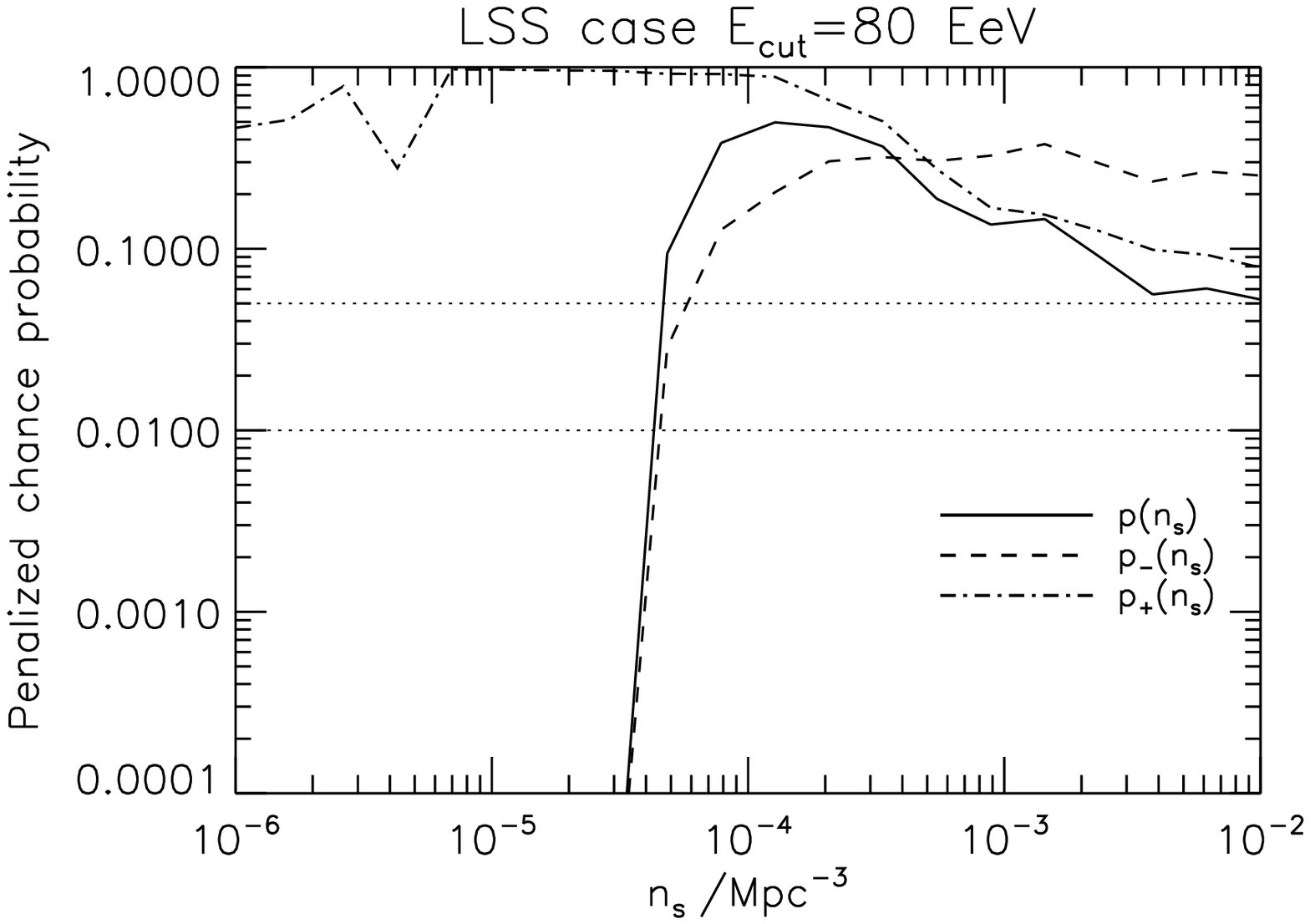,width=0.47\textwidth}\\
\end{tabular}
\end{center}
\vspace{0pc} \caption{Penalized chance probabilities $p_{-}(n_s)$,
$p_{+}(n_s)$ and $p(n_s)$, for  $\Ecut=60\,$EeV (top panels) and
$\Ecut=80\,$EeV (bottom panels). The left column reports the case
for uniformly distributed sources, the right panel for sources
following LSS with the bias of the PSCz Galaxy catalogue. Also shown
is the 95\% and 99\% confidence level.}\label{fig:6}
\end{figure*}

The usual way to circumvent the problem is to use the Monte Carlo to
calculate the chance probability to observe stronger clustering than
in the data. Given the problem at hand we slightly generalize the
method defining two functions: the chance probability to observe
stronger ($P_+(\theta|X)$) or weaker ($P_-(\theta|X)$) clustering at
the angular scale $\theta$ than in the data given by,
\be
 P_{\pm}(\theta|X) =
 \frac{1}{M}\sum_{i=1}^M \Theta[\pm(C_i(\theta|X)-C_{\ast}(\theta))]
 \,.\label{Pdelta}
\ee
The minimum of the chance probability can then be used as a global
estimator for the agreement of the hypothesis with the data. In our
case with a scan of $P_\pm(\theta|X)$ over $\theta$ we obtain the
two minima $P_-(X)$ for the minimum of $P_-(\theta|X)$ and $P_+(X)$
for the minimum of $P_+(\theta|X)$, respectively. The simplest way
to combine these two estimators is then to use the product
$P(X)=P_-(X)P_+(X)$.

However, a drawback of this method is that neither the quantity
$P(X)$ or the single $P_\pm(X)$ are truly probabilities. The scan
over the angular scales $\theta$, in fact, introduces a bias which
need to be corrected with the use of a penalty factor
\cite{Tinyakov:2001nr,Tinyakov:2003bi,Finley:2003ur}. More precisely, to obtain correct
probabilities the identical procedure as described above needs to be
performed with many simulated data sets. Counting how often smaller
values of $P_\pm(X)$ and $P(X)$ are obtained by chance with respect
to the case of the data, provides true penalized chance
probabilities $p_{+}(X)$, $p_{-}(X)$, and $p(X)$.

The use of the chance probability tool has often generated some
confusion in the past. An emblematic case is the significance of the
small scale clustering in the AGASA data for which very different
estimates ranging from $\sim 10^{-6}$ to $\sim 10^{-2}$ has been
reported in various studies (see for example Ref. \cite{auto}) depending
on the use or not of the penalty  and on different a priori choices
of the angular and energy scale of reference (see
\cite{Finley:2003ur} for a detailed account). The effect of the scan
can then be quite relevant and a major point in the following is
that the effect of the penalty is correctly taken into account when
quoting the constraints on $n_s$ and the constraints are further
compared with the case of an a priori choice of the relevant angular
scale.

Note, anyway, that the penalty calculation can be avoided if a
particular angular scale is chosen \emph{a priori} and the values of the chance probability at this scale are
employed. However, the scan over all angles avoids possible bias, in
contrast to the choice of a single angular scale, which introduces
some theoretical prejudice even if this choice may be physically
motivated. In the case at hand, it is unclear if this should be
dominated by: The $\sim 1^\circ$ angular resolution of the detector,
as in~\cite{Takami:2008rv} which implicitly assumes negligible
magnetic field deflections; by a $\sim 3^\circ$ scale, as suggested
by the cross-correlation with active galactic nuclei (AGN) revealed
by Auger~\cite{Cronin:2007zz}\footnote{Note that, for un-correlated
deflections, the window size to use for autocorrelation studies
would be $\sqrt{2}\times 3.1^\circ\sim 4.3^\circ$; actually, since
deflections from the source are correlated and the energies of
events similar, the relative deflections for a single chemical
species would be likely $\alt 3^\circ$.}; or yet some other scale,
as the $6^\circ$ separation considered in~\cite{PAO}. To emphasize
this dependence, we summarize the results of  this kind of
``single-bin'' analyses in table~\ref{tab:1} and compare them with
the ones obtained with the global method, reported in the last
column (see next Section). Note, in particular, how the 6$^\circ$
bin chosen in~\cite{PAO} leads systematically to an overly stringent
bound.
%%%%%%%%%%%%%%%%%%%%%%%%%%%%%%%%%%%%%%%%%%%%%%%%%%%%%%%%%%%%%%%%%%%%%%%%%%%%
\section{Interpretation}\label{int}
%%%%%%%%%%%%%%%%%%%%%%%%%%%%%%%%%%%%%%%%%%%%%%%%%%%%%%%%%%%%%%%%%%%%%%%%%%%%

\begin{figure}[!b]
\begin{center}
\begin{tabular}{c}
%\hspace{2pc}
\epsfig{file=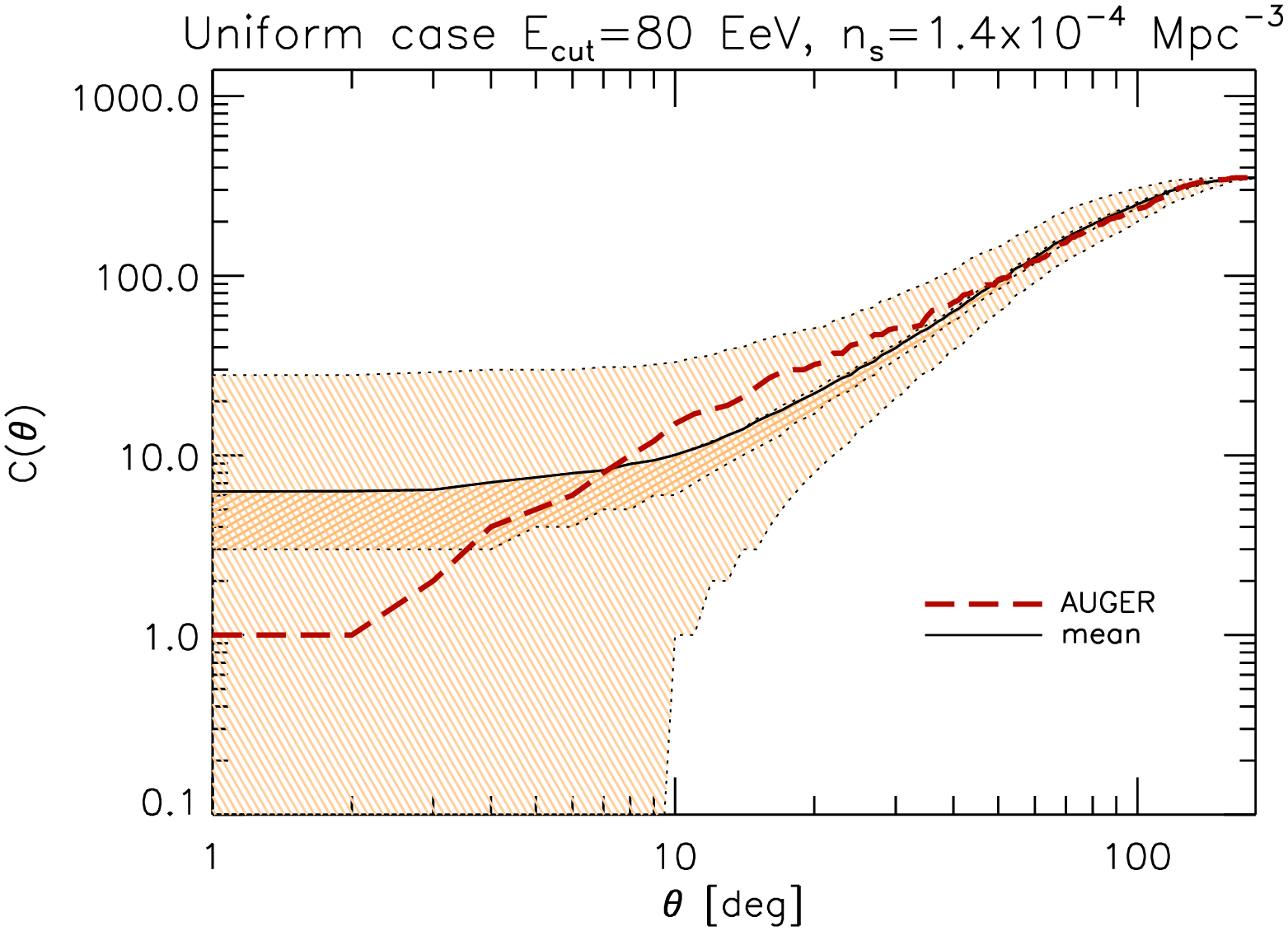,width=0.51\textwidth}\\
%\hspace{2pc}
\epsfig{file=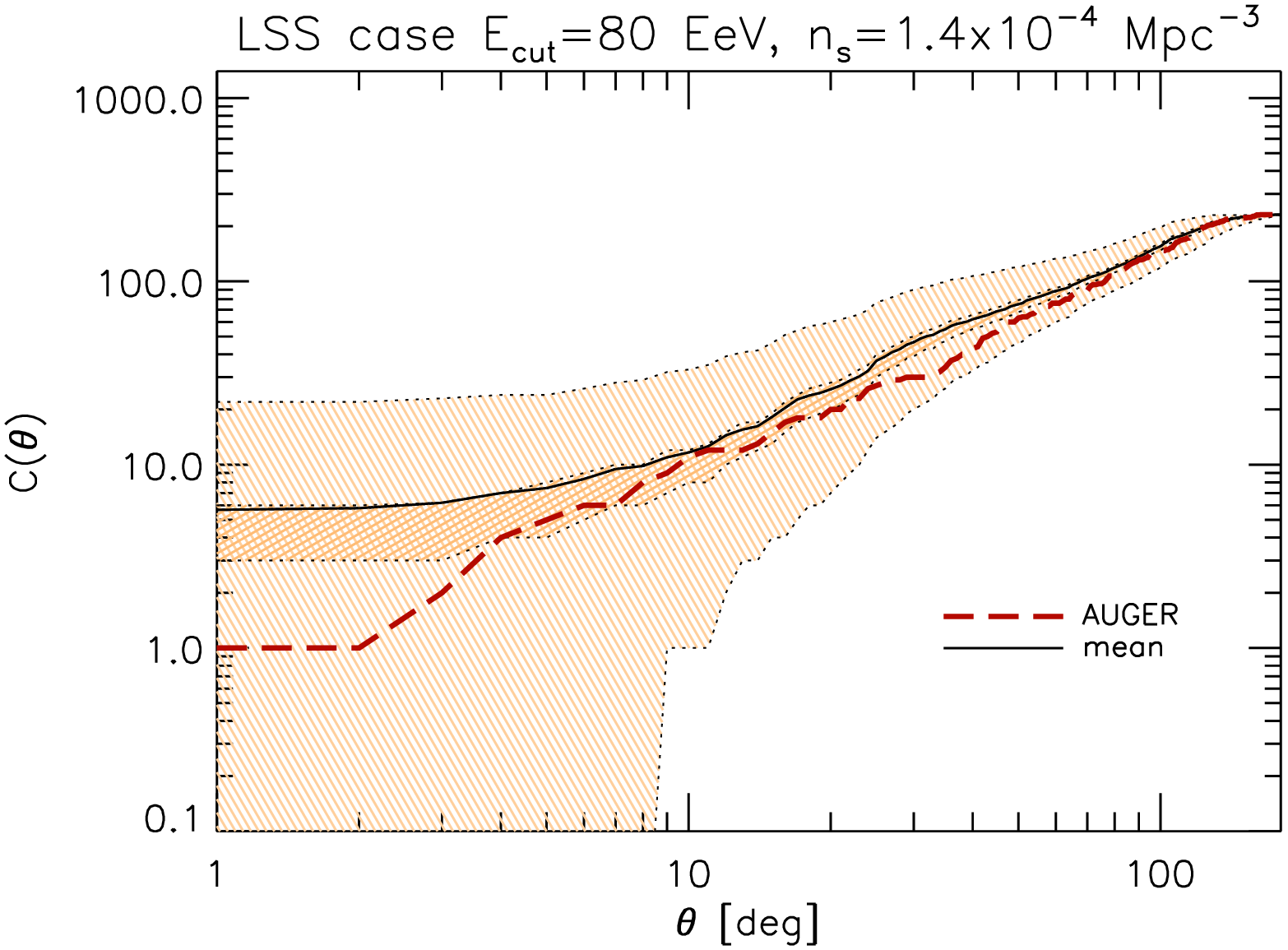,width=0.51\textwidth}
\end{tabular}
\end{center}
\vspace{0pc} \caption{Cumulative autocorrelation function for
$N_{\rm Auger}=$27  events compared with
the best fit model for the uniform (top) and LSS (bottom) cases. In the
latter case, only 22 events survive outside the PSCz mask.
Both plots assume with $\Ecut=80\,$EeV. The mean model autocorrelation is shown
together with the 1$\sigma$ and $3\sigma$ regions.}\label{fig:2}
\end{figure}

\begin{table*}[!t]
\begin{center}
\footnotesize{
\begin{tabular}{|c|c|c|c|c||c|}
\hline $n_s/10^{-4}$Mpc$^{-3}$, \,  $\theta_1$= & $6^\circ$ &
$3\sqrt{2}^\circ$ & $3^\circ$ & 1$^{\circ}$ & global \\ \hline
 \hline LSS (80)&        $\; 1.3^{+5.7}_{-1.0} \;$    & $\; 2.0^{+8.0}_{-1.6} \;$    & $\; 2.0^{+\infty}_{-1.4} \;$   & $\; 5.0^{+\infty}_{-4.2} \;$  & $\; 1.3^{+100}_{-0.8} \;$\\
 \hline Uniform (80)&    $\; 0.8^{+0.8}_{-0.5} \;$    & $\; 1.2^{+0.8}_{-0.8} \;$    & $\; 2.5^{+\infty}_{-1.8} \;$      & $\; 5.0^{+\infty}_{-4.2} \;$  & $\; 1.4^{+1.4}_{-0.7} \;$\\
 \hline LSS (60) &       $\; 0.3^{+1.7}_{-0.27} \;$   & $\; 0.3^{+2.7}_{-0.18} \;$   & $\; 0.7^{+100}_{-0.5} \;$     & $\; 1.5^{+100}_{-1.25} \;$     & $\; 0.8^{+19}_{-0.6} \;$\\
 \hline Uniform (60) &   $\; 0.2^{+1.0}_{-0.12} \;$   & $\; 0.3^{+1.7}_{-0.2} \;$   & $\; 0.8^{+70}_{-0.63} \;$       & $\; 1.0^{+\infty}_{-0.8} \;$     & $\; 0.5^{+0.5}_{-0.2} \;$\\
\hline
\end{tabular}  }
\end{center}
\caption{\label{tab:1} The estimated number density of sources (at
95\% confidence level) under different assumptions, using only the
first bin information with different sizes, and compared with the
global method.}
%\vskip0.3cm
\end{table*}

In Fig.~\ref{fig:6}, we report the results for the quantities $p_i(X)$
defined above, where $X\equiv\{n_s, E_{\rm cut},\kappa\}$, the
latter two being in our case two-valued discrete variables. Because
the number of CRs usable for this analysis is still small, all four
hypotheses are compatible with the data at the $2\sigma$ level for
some range of $n_s$ values. Yet, several interesting conclusions
can already be drawn. The best fit is achieved for sources following
the PSCz distribution and a
source density $n_s\simeq 1\times 10^{-4}$/Mpc$^3$. Also,
independently of $E_{\rm cut}$, both the penalized probability and
the range of $n_s$ with compatibility at 95\% C.L. are larger for
the LSS model than for the uniform case. Reversing the argument, as
can be read more quantitatively in table~\ref{tab:1}, we can see
that the constraints are generally stronger for the uniform cases
with respect to the LSS ones, but this is achieved only at the
expense of a worse general best fit. The fact that the LSS models fit
better the data is not surprising: Most of the Auger event are aligned
along the local overdensity known as the Supergalactic plane which is
suitably reproduced with the use of the PSCz catalogue within our
LSS scenario.

The case of a uniform distribution of ``infinitely many" sources
($n_s\to \infty$ each with an infinitesimal luminosity), is excluded
for both energy cuts at the 95\% C.L.: The upper bound is $n_s\alt
(1\div 3)\times 10^{-4}\,$Mpc$^{-3}$. This is another way to say
that {\it the Auger data are inconsistent with a structureless UHECR
sky,  independently of the use of a catalogue and of a
pre-determined angular scale for the search.\/} This is, in our
opinion, an important milestone in the development of UHECR
astronomy. While the best fit point for $n_s$ is approximately a
factor 10 higher than found in earlier studies using AGASA data
above $E_{\rm cut}> 40\,$EeV (in the AGASA energy
scale)~\cite{ns1,blasi,ns2}, the shape of the chance probability $p(n_s)$
agrees: For low values of $n_s$, $p(n_s)$ is a steeply decreasing
function of $n_s$, since the probability to observe multiplets from
the same source increases fast. In particular, the radius within
which 70\% of all observed UHECRs with energy above $E_{\rm
cut}=80\,$EeV are produced is $R\approx 60$
Mpc~\cite{Cuoco:2005yd}\footnote{The quoted value depends on the use
of the continuous-energy loss approximation, the actual value
increasing to \mbox{$\approx$ 70 Mpc} due to the stochastic nature
of the photo-pion production and to \mbox{$\approx$ 100 Mpc} further
considering a 20\% energy resolution~\cite{res}. For the estimate of
$n_s$, however,  we do not use the concept of horizon size
explicitly, which is here introduced only for illustration.}. As a
result, the number of sources within this radius becomes less than
the number of observed CRs events for densities smaller than
$n_s\approx10^{-5}$. Such a scenario would require large deflections
(and probably nuclei primary) and thus contradicts our assumptions.
On the other hand, $p(n_s)$ decreases relatively slowly for high
densities and only weak constraints can be obtained with the current
data set  for the maximally allowed value of $n_s$. Since both an
increase of $E_{\rm cut}$ and of the bias of the sources leads to a
decrease of the effective number of sources inside the GZK volume,
large values of $n_s$ have the strongest constraint in the case of
uniformly distributed sources and $E_{\rm cut}=60$\,EeV (left, top
panel) and weakest for sources following the LSS and $E_{\rm
cut}=80$\,EeV (right, bottom panel).

In Fig.~\ref{fig:2} we show the model autocorrelation function with
$1\sigma$ and $3\sigma$ shaded regions for the best fit uniform and
LSS model for $\Ecut=80\,$EeV, both corresponding to $n_s\approx
1.4\times 10^{-4}\,$Mpc$^{-3}$, together with the data. At small
angular scales, $\theta\lsim 3^\circ$, the data show a deficit of
clusters compared to the expectation for the the best fit density
from the global analysis. This ``tension'' is qualitatively present
in most models fitting the data. Within our assumptions, this
deficit is explained in a natural way by (relative) deflections of
this size in magnetic fields. This value is comparable with the
$3.2^\circ$ found in Ref.~\cite{PAO} that optimizes the correlations
of the same data set with AGNs. The absence of small-scale clusters
is also responsible for the shift in the best fit value for $n_s$
compared to old analyses using the AGASA data. The result thus shows
that, intriguingly, also the autocorrelation function can be
employed as a sensitive tool for magnetic field studies. Clearly,
however, a more detailed study needs to be complemented with a model
of the intervening magnetic fields while, likely, more statistics is
required to derive significant constraints.

Our result is in contrast with the one of \cite{Takami:2008rv} which
instead report a small scale clustering in the Auger data. Notice
however that, with respect to our work, the authors of
\cite{Takami:2008rv}, although including an explicit treatment of
galactic and extragalactic magnetic fields, make use of the
non-cumulative autocorrelation functions and do not take into
account penalties for the scan over the angular scale. Their claim
of small scale clustering within $1^\circ$ is thus likely to
disappear if the angular scan is taken into account and the
comparison is made properly with respect to the a model effectively
fitting the data. Thus, as already noted in \cite{Harari:2004py},
the interpretation of small scale clustering could be misleading if
it is not defined with respect to a proper model of the distribution
of sources. Apart from this point, however, the constraints we obtain
on $n_s$ are roughly in agreement with
\cite{Takami:2008rv}. This is mainly due to the very low statistics
available at present which is not still sensitive to the exact
analysis method employed. We stress however, as noted in \cite{I},
that already with a statistic of $\sim$3 times the present one a
formally correct analysis will become crucial to avoid biased
results.

Figure~\ref{fig:2} also clarifies why the LSS case gives a better fit
to the data: As can be seen, the LSS best fit model fits nicely the
data, basically within $1\sigma$ over all the angular scales, while
the best fit case for  uniform sources shows a $\sim2\sigma$ deficit
in the broad
range $4^\circ$--$30^\circ$. A lower $n_s$ (i.e.\ an higher
clustering) would not help because it would give much more pairs
than the data in the  $1^\circ$--$4^\circ$ range. Thus, at the end,
even with the best possible compromise the agreement with the data
is only at the 20\% level for the best fit (or, equivalently, the
uniform model is excluded at the 80\% C.L.).

\begin{figure}[!t]
\begin{center}
\begin{tabular}{c}
\hspace{1pc}
\epsfig{file=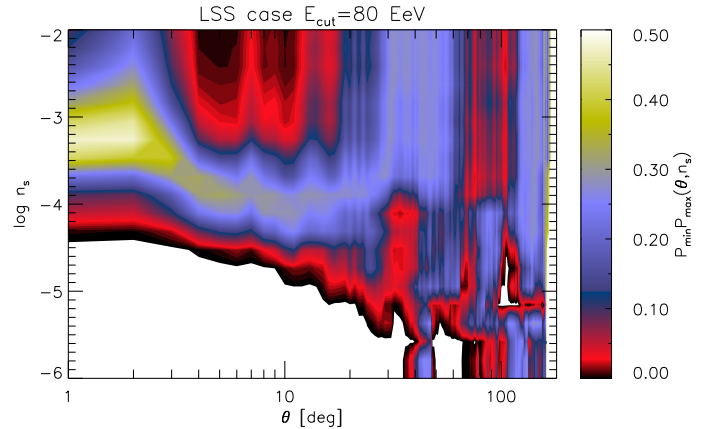,width=0.45\textwidth}
\end{tabular}
\end{center}
\vspace{0pc} \caption{Contours of equal chance probability
$p(\theta|n_s)$ for the LSS case and $\Ecut=80\,$EeV. \label{fig:3}}
\end{figure}

We summarize in table \ref{tab:1} the list of best fit $n_s$ with
95\% error bars for the four cases considered and for different
choices of the angular scale or for the global autocorrelation
analysis. The crucial point to notice here is that the derived $n_s$
intervals are sensibly biased with respect to each other for
different choices of the a priori angular scale. Thus, \emph{the
choice of the angular scale crucially affects the result and should
be avoided unless the given scale  has some strong physical
motivation}. This problem is of course avoided employing a global
comparison. The choice of a single angular scale also affects the
error bars which can be both larger or smaller with respect to the
global case. This can easily understood from Fig.~\ref{fig:3} where
we can see that the choice $4^\circ$--$15^\circ$ is optimal  because
excess of clustering is observed for low $n_s$ while a deficit is
present for high $n_s$ giving thus the tightest constraints. Again,
however, a choice in this range is not a priori motivated so that
from the global analysis we get somewhat larger error bars properly
taking into account all the angular scales.

Although it is difficult to draw strong conclusions on the nature of
the UHECR sources at this moment, it is worth noting  that  X-ray
selected AGNs with X-ray luminosity $L > 10^{43}\,$erg/s naturally
fall in the range $n_{\rm AGN}=(1\div 5)\times 10^{-5}\,{\rm
Mpc}^{-3}$~\citep{Steffen:2003uv}. at the same time, for AGNs with
densities in the range $(10^{-5}\div10^{-4})\,{\rm Mpc}^{-3}$, the
required luminosity is of the order ${\cal L}/n_{\rm AGN}\sim
(10^{40}\div 10^{41})\,$erg/s in UHECRs above $\sim 60\,$EeV. So,
simple energetics arguments are consistent with the inferred values
for the density; on the other hand, acceleration efficiency (see
e.g.~\citep{PT08}) tends  to favor some subclasses of objects and
thus a lower inferred density of sources, yet typically within the
95\% C.L. range inferred for $n_s$. Another effect which might
reconcile the small tension is that the sources can be
bursting/transient or beamed. In this case the luminosity
requirement is softened \cite[see e.g.~the model proposed
by][]{Farrar:2008ex}. However, the effectively visible number
density of sources decreases, too, unless some isotropization takes
place after acceleration but close to the source \citep[see][for
more details]{Sigl:2008fr}. Of course, these are very preliminary
considerations based on a limited number of data. For example, these
arguments might be significantly modified when accounting for a more
realistic luminosity function. The effects of extra-galactic
magnetic fields and the consequent UHECRs time delays can also be
quite relevant, although at present our knowledge of EGMFs is
affected by large uncertainties and needs to be better understood
(see \cite{Sigl:2008fr,dgst,Kotera:2008ae,Ryu:2008hi}). Also, in
presence of a heavy-nuclei component and/or of extragalactic
magnetic fields, a significant fraction of events might be
associated with the nearest AGN Cen A \citep[see
e.g.][]{gorbunov08}. Interestingly, signatures in the  gamma and
neutrino bands are expected to help disentangling among many
scenarios
(see~\cite{Sigl:2008fr,Cuoco:2007qd,Kachelriess:2008qx,Becker:2008nf,Halzen:2008vz,Hardcastle:2008jw}),
enlarging the realm of multi-messenger astronomy.

%%%%%%%%%%%%%%%%%%%%%%%%%%%%%%%%%%%%%%%%%%%%%%%%%%%%%%%%%%%%%%%%%%%%%%%%%%%%
\section{Discussion on some simplifying assumptions}\label{disc}
%%%%%%%%%%%%%%%%%%%%%%%%%%%%%%%%%%%%%%%%%%%%%%%%%%%%%%%%%%%%%%%%%%%%%%%%%%%%

\subsection{The role of the energy cut}
The energy-cut used by the PAO to produce the sample used in the
present analysis was chosen in order to maximize the
cross-correlation signal, see~\cite{Abraham:2007rj,PAO} for details.
One may wonder how this selection affects the conclusions of  our
present work. In principle, the optimal energy cut for the
autocorrelation signal and the cross-correlation signal with a given
catalogue differ one from another, although in the case of a common
physical origin one does expect some correlation between them. In
particular, it seems reasonable that the optimal cut for a global
autocorrelation might reside at a  lower energy, since the
small-scale displacement from putative sources is not as relevant to
the signal as for cross-correlations, and the larger statistics
helps. Lacking a direct access to the data, it is hard to
estimate quantitatively how large a bias is introduced by focusing
on the sample of public available events.  From the Fig. 2 presented
in~\cite{Mollerach:2007vb}, however, one can draw two qualitative
conclusions: i) that a correlation between the two optimal cuts in
energy is indeed present; ii) that a slightly different cut, in the
range $40\lsim E_{\rm cut}/{\rm EeV}\lsim 60$, should still lead to an appreciable clustering
of the events.

\begin{figure*}[!htbp]
\begin{center}
\begin{tabular}{cc}
\epsfig{file=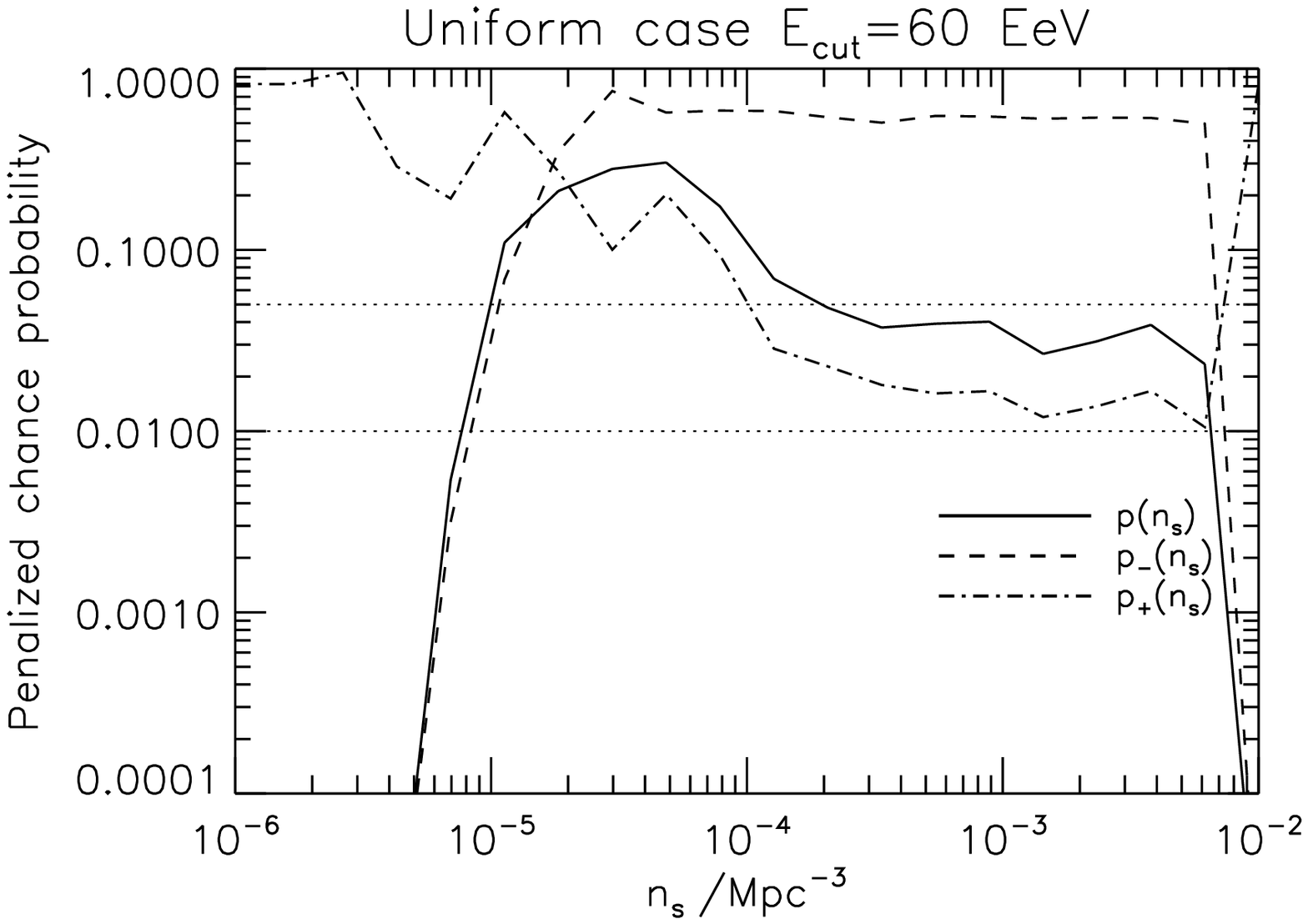,width=0.47\textwidth}
&
\epsfig{file=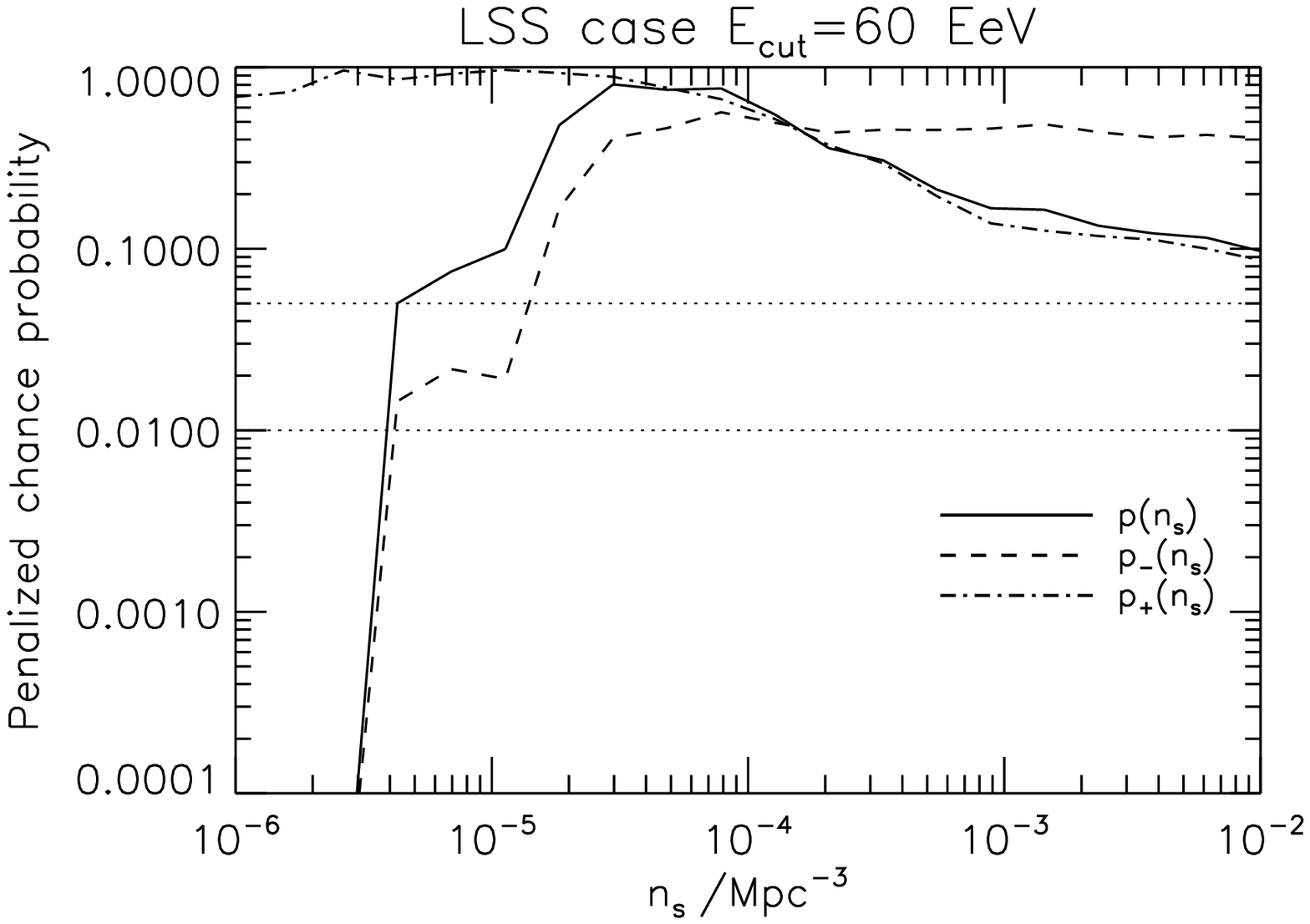,width=0.47\textwidth}\\
\end{tabular}
\begin{tabular}{cc}
\epsfig{file=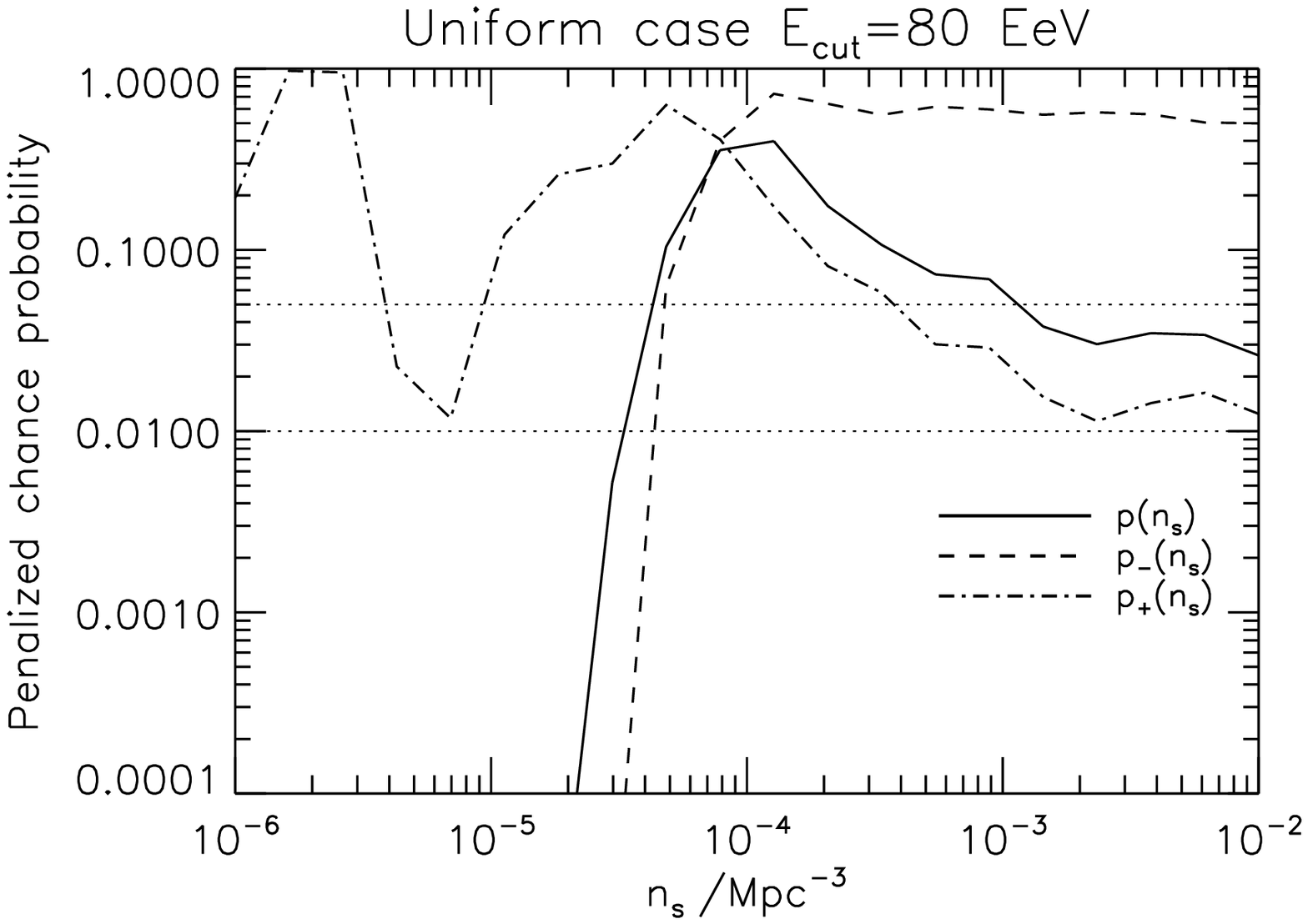,width=0.47\textwidth}
&
\epsfig{file=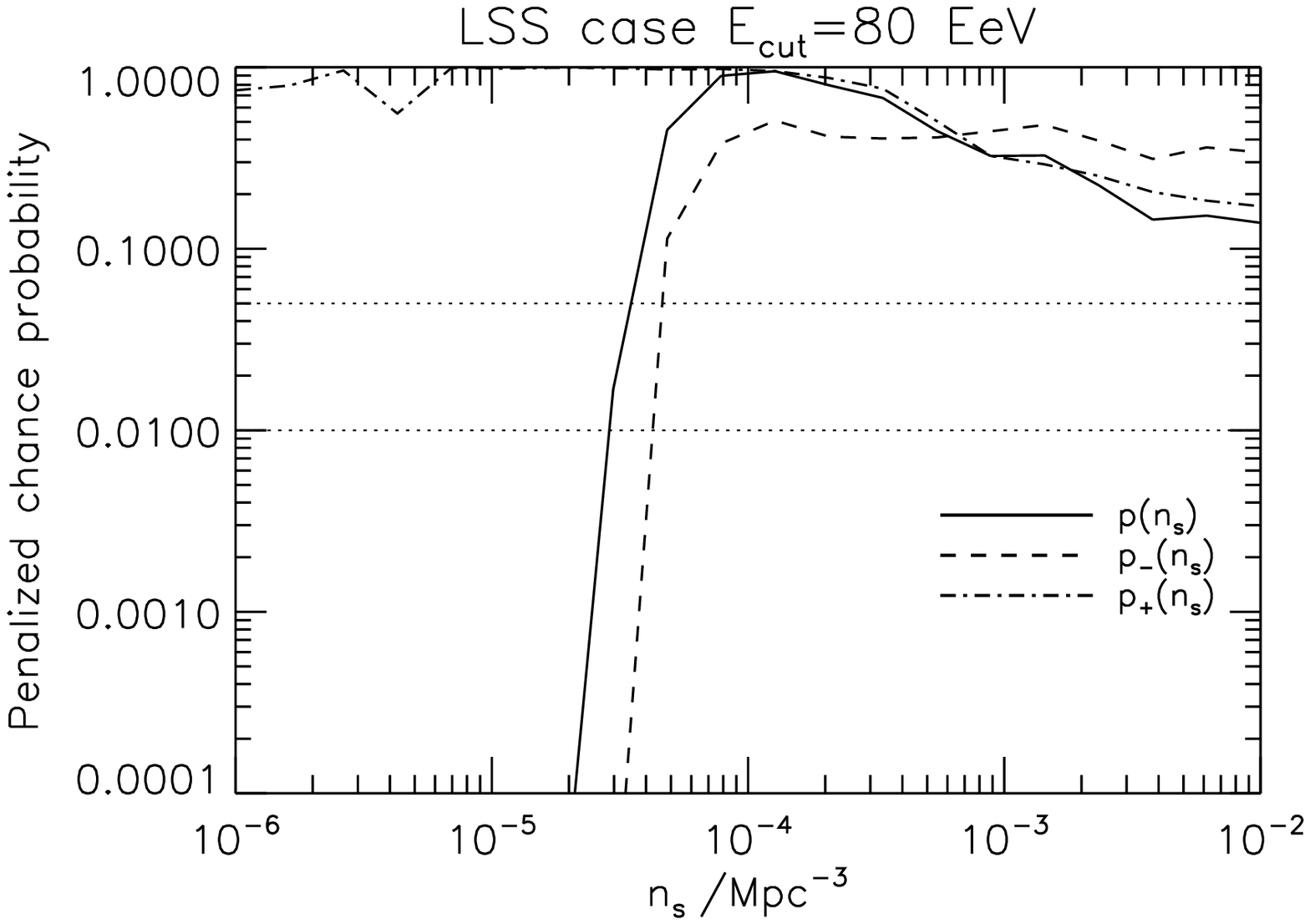,width=0.47\textwidth}\\
\end{tabular}
\end{center}
\vspace{0pc} \caption{ Penalized chance probabilities $p_{-}(n_s)$,
$p_{+}(n_s)$ and $p(n_s)$, for  $\Ecut=60\,$EeV (top panels) and
$\Ecut=80\,$EeV (bottom panels) taking into account the effect of
the energy scan. The left column reports the case for uniformly
distributed sources, the right panel for sources following LSS with
the bias of the PSCz Galaxy catalogue. Also shown is the 95\% and
99\% confidence level.}\label{fig:energyscan}
\end{figure*}

We checked that this is indeed the case by performing the same
analysis  as before, but now adding a further scan  over the range of
accessible energy-cuts, i.e. any value $E>57\,$EeV. We plot the
results in Fig.~\ref{fig:energyscan}\footnote{Some ripples visible in Fig.~\ref{fig:energyscan} are due to the
relatively low number of Montecarlo: the scan to account for the further penalty factor
is computationally quite expensive and given the partial nature of the answer there
is no motivation to refine the results further.}. One can note that
in all cases the best fit improves and the constraints on $n_s$ worsen, as
expected given the further penalty due to the energy scan.
Yet, most qualitative features described in the previous
section stay the same: for example, the LSS model is still
preferred over a uniform one.
It should be also said that if the true minimum of the chance
probability is below $E=57$ EeV and thus not included in the scan,
then these constraints are ``over-penalized'' and thus looser than
necessary. At the moment, it is impossible to draw more quantitative
conclusions,   since our scan suggests that it is likely that the
optimal cut for the autocorrelation function is below $E=57$ EeV, a
range for which the events are not publicly available.

%----------
\subsection{Assumptions on the chemical composition}\label{chemadd}
%----------
In this article, we assumed dominant proton primary as a basic
working hypothesis. It is worth noting, however,  that the
experimental situation on the chemical composition at UHE is far
from  settled:  while anisotropy data point to a relatively light
composition, the results of the fluorescence detector of the Pierre
Auger Collaboration favor a significant fraction of heavy
nuclei~\cite{Unger:2007mc}. Yet, for the interpretation of these
results one must rely on simulations employing hadronic interaction
models. These are not based on a first-principle theory, rather on
models calibrated on ``low-energy''  collider data, then {\it
extrapolated} about two orders of magnitude beyond the center of
mass energies experimentally probed.

A proper quantitative assessment on how our conclusions vary in  a
mixed composition scenario goes beyond the purpose of the present
paper. Yet, at a qualitative level, we can note that several effects
would come into play. First of all,  in the unrealistic case where
one could forget about magnetic deflections, the major effect would
be a reduction of the energy-loss horizon (but for iron, whose
horizon is similar to the proton one). This should {\it enhance}
the anisotropy pattern, due to the prominence of nearby
accelerators.

When including (the poorly known) magnetic fields, two additional
effects are relevant: i) for a given energy, the higher the charge
the larger the deflection and hence the  loss of information at
{\it small} angular separations. Quantifying how large is this scale
is a difficult task. We note that protons of these energies in the
sole Galactic field likely suffer a few degrees absolute
deflections, see~\cite{Kachelriess:2005qm}, implying a degree-scale
smoothing in the relative deflections important for the 2pcf. This
is comparable with the angular resolution of the PAO: in this
optimal case the whole information in the 2pcf starting at
$\theta_{\rm min}\sim 1^\circ$ could be used. However, a different
Galactic field model (especially towards the Galactic Center), the
presence of heavy nuclei, and/or significant extragalactic magnetic
fields can easily lift $\theta_{\rm min}$ by one order of magnitude
or more. ii) the other effect is that the real path-length of the
nucleus would be longer than the distance to the source: thus, the
distance of accessible sources would be even shorter than estimated
from energy-loss considerations. This is particularly relevant if
the nucleus spends a lot of time in a magnetized region surrounding
the accelerator (e.g. a magnetized cluster in which it is immersed)
before escaping in the Intergalactic Medium.

Finally, if the maximal energy of acceleration of different  species
of nuclei fall by unfortunate coincidence in the same region
expected for the GZK feature, slightly different energy cuts in the
data (as well as statistical and systematic errors on the energy
scale) might significantly change the expected pattern of
anisotropies. The same happens if the proportions of different
nuclei accelerated at the source change as a function of the energy.
This is in principle a possibility, especially if different classes
of objects contribute to the events at slightly different energies.

The general pessimistic conclusion is that if several of  the above
effects are relevant (or perhaps a single one is {\it dominant}) the
capability of performing UHECR astronomy would be greatly reduced.
While one still expects indications for anisotropies, inverting the
problem and  inferring the source/propagation medium properties
would require a much larger statistics (especially in the trans-GZK
region): disentangling the different effects is in fact a formidable
task.

To provide a glimpse of how some of the  above effects alter the
reconstruction of $n_s$ (our main topic here), in Table~\ref{tab:2}
we report how the constraint on $n_s$ degrades as a function of a
``smoothing angle'' $\theta_{\rm min}$, below which we assume that
the 2pcf information is completely lost.  The main trend is that, if
the smallest angular scales are neglected, it is easier to find
parameter configuration fitting the data and, correspondingly, the
allowed range for $n_s$ widens. This had to be expected, given the
shape of the correlation functions shown in Figs. 2-3.   In
particular we find that with $\theta_{\rm min}=3^\circ$ we obtain
almost the same results as in the global case. The case $\theta_{\rm
min}=10^\circ$ still places useful constraints especially for the
$\Ecut=80\,$EeV case, while finally using only the information above
$\theta_{\rm min}=30^\circ$ basically no constraints on $n_s$ are
obtained. Note however that relative deflection angles of that sort would
imply overall deflections even larger, seriously questioning the perspectives of present
instruments to perform some form of UHECR astronomy.

\begin{table*}[!t]
\begin{center}
\footnotesize{
\begin{tabular}{|c|c|c|c||c|}
\hline $n_s/10^{-4}$Mpc$^{-3}$, \,  $\theta_{\rm min}$= & $3^\circ$ &
$10^\circ$ & $30^\circ$ & global \\ \hline
 \hline LSS (80)&   $\; 1.0^{+100}_{-0.8}  \;$      &  $\; 0.5^{+30}_{-0.4}    \;$&  $\; 0.5^{+\infty}_{-0.4} \; $ & $\; 1.3^{+100}_{-0.8} \;$\\
 \hline Uniform (80)& $\;0.8^{+1.5}_{-0.6} \;$      &  $\; 0.3^{+1.7}_{-0.2}   \;$&  $\; 0.3^{+\infty}_{-0.2} \; $ & $\; 1.4^{+1.4}_{-0.7} \;$\\
 \hline LSS (60) &   $\; 0.3^{+20}_{-0.28}   \;$    &  $\; 0.1^{+5}_{-0.09}    \;$&   n.c.                   & $\; 0.8^{+19}_{-0.6} \;$\\
 \hline Uniform (60) & $\; 0.2^{+0.8}_{-0.12} \;$   &  $\; 0.1^{+0.9}_{-0.085} \;$&   n.c.                   & $\; 0.5^{+0.5}_{-0.2} \;$\\
\hline
\end{tabular}  }
\end{center}
\caption{\label{tab:2}  The estimated number density of sources (at
95\% confidence level) under different assumptions on the minimum
angle above which the 2pcf information is preserved, $\theta_{\rm
min}$. n.c. stands for ``no constraints''.}
\end{table*}
%-----

%%%%%%%%%%%%%%%%%%%%%%%%%%%%%%%%%%%%%%%%%%%%%%%%%%%%%%%%%%%%%%%%%%%%%%%%%%%%
\section{Summary}\label{sum}
%%%%%%%%%%%%%%%%%%%%%%%%%%%%%%%%%%%%%%%%%%%%%%%%%%%%%%%%%%%%%%%%%%%%%%%%%%%%
We used the first UHE data released by Auger  to perform a {\it
global} autocorrelation study of UHECR arrival directions, assuming
proton primaries. The major advantage of our tool is that no biases
are introduced by the \emph{a priori} choice of a single angular
scale. The main observable we have focused on is the number density
$n_s$  of ultrahigh energy cosmic ray (UHECR) sources. While the
global analysis does not bias $n_s$ by  what is the theoretical
prior of the ``relevant'' angular scale, still it is important to
establish how the extraction of the allowed range of $n_s$ from the
data depends on a number of other effects, an issue often overlooked
in the literature. In particular, here we discussed the systematic
energy scale uncertainty and of the bias of UHECR sources with
respect to Large Scale Structures. As a first attempt to extract
some information from the data, we compared four hypotheses: a
structured universe (following the PSCz catalogue) and an isotropic
case, each for two possible values for the absolute energy scale of
the PAO experiment. The density $n_s$ is the free continuous
parameter in terms of which constraints have been analyzed.

Not surprisingly, we find that the number  of CRs usable for this
analysis is still small and all four hypotheses are compatible with
the data at the $2\sigma$ level for some range of $n_s$. Yet,
several interesting observations can be tentatively drawn. The best
fit is achieved for sources following the matter tracer
distribution, and a source density $n_s\simeq 1\times
10^{-4}$/Mpc$^3$. It is interesting to note that the data show some
preference (although not significant, yet) for a structured
universe: other recent statistical studies,
like~\cite{Kashti:2008bw,Koers:2008ba}, qualitatively agree in that
respect. Also, there is indication that the case of a uniform
distribution of ``infinitely many" sources ($n_s\to \infty$ each
with an infinitesimal luminosity), is excluded for both energy cuts
at the 95\% C.L.: The upper bound is $n_s\alt (1\div 3)\times
10^{-4}\,$Mpc$^{-3}$. This is another way to say that early Auger
data suggest that data are poorly consistent with a structureless
UHECR sky,  independently of the use of a catalogue and of a
pre-determined angular scale for the search.

Compared to a benchmark number density of proton sources $n_s\simeq
1\times 10^{-4}/$Mpc$^3$,  a factor $\sim 2$ lower densities are
preferred if the current Auger energy scale is correct, a factor
$\sim 2$ higher value if it is underestimated as
required by the {\it dip} model. Including the finite energy
resolution of  the Auger experiment into the analysis will reduce
further the best fit value for $n_s$. The width of the allowed
region is dominated at present by the statistical error due to the
small number of events. Nominally, approximately a three times
larger sample is needed to reduce the Poisson error below the
typical differences between source candidates. Once that level of
statistics is reached, other effects will provide the main source of
error, a major one being the systematics on the energy scale as we
illustrated here. In the future, other effects such as the
systematic and statistical errors in the energy determination of
UHECR events and the stochastic nature of the photo-pion process
need to be included to correctly determine the best fit value of
$n_s$.  Even considering the above limitations, preliminary
conclusions are the following: First, the fit generally improves for
sources following the LSS compared to uniformly distributed sources;
qualitatively, for AGNs which are known to be even more
overdensity-biased than the LSS (see e.g. discussion in \citep{I}),
the agreement is expected to improve even further. In particular,
the case of an uniform distribution  of ``infinitely many" sources,
i.e.\ a structureless UHECR sky, is excluded at 99\% C.L. Second,
the absence of clustering on scales smaller than a few degrees is
most  easily understood as the effect of magnetic smearing of
comparable size. This, intriguingly, suggests that autocorrelation
studies can be employed as a complementary tool to study galactic
(and extragalactic) magnetic fields.
 Conclusions about the source
scenario are complicated by the the limited knowledge of the EGMFs
and the presence of beaming and/or bursting effects which are
difficult to disentangle with the use of the autocorrelation alone.
Hence, the true source density could be somewhat lower that the best
fit value found: The energy scale error alone shifts the best fit
value by a factor two or three; luminosity function effects are
likely relevant, too. The scarce statistics at the moment is a
serious limiting factor to constrain more realistic models, but with
a greater exposure of currently existing instruments  and the
multi-messenger combination of gamma-ray and neutrino data, some of
these issues will probably be addressed and solved in the near
future.

For a significant contamination of nuclei in the sample,  a much
more complicated analysis is needed,  since many more variables
enter the game. Qualitatively, one can expect that although future
data might allow to disentangle a proton-dominated sample from a
more complicated picture, inferring source properties and
disentangling them from magnetic effects (in a few words, performing
UHECR astronomy) should wait for a major jump in exposure, perhaps
beyond the capabilities of currently planned  instruments. Similar
considerations apply unfortunately to the constraints to the cross
sections of UHECRs as well.

%%%%%%%%%%%%%%%%%%%%%%%%%%%%%%%%%%%%%%%%%%%%%%%%%%%%%%%%%%%%%%%%%%%%%%%%%%%%
\section*{Acknowledgments}
%%%%%%%%%%%%%%%%%%%%%%%%%%%%%%%%%%%%%%%%%%%%%%%%%%%%%%%%%%%%%%%%%%%%%%%%%%%%
We are grateful to G\"unter Sigl for useful discussions. M.K.\ thanks
the Max-Planck-Institut f\"ur Physik in Munich for hospitality and
support.   P.S. is supported by the US
Department of Energy and by NASA grant NAG5-10842. Fermilab is
operated by Fermi Research Alliance, LLC under Contract
No.~DE-AC02-07CH11359 with the United States Department of Energy.

%%%%%%%%%%%%%%%%%%%%%%%%%%%%%%%%%%%%%%%%%%%%%%%%%%%%%%%%%%%%%%%%%%%%%%%
%\section*{References}
%%%%%%%%%%%%%%%%%%%%%%%%%%%%%%%%%%%%%%%%%%%%%%%%%%%%%%%%%%%%%%%%%%%%%%%

%%%%%%%%%%%%%%%%%%%%%%%%%%%%%%%%%%%%%%%%%%%%%%%%%%%%%%%%%%%%%%%%%%%%%%%%%%%%

\end{document}